\documentclass[prd,aps,twocolumn,a4paper]{revtex4}

\def\p{\partial}
\def\Lie{{\cal L}}
\def\br{\nonumber \\ &&}
\def\scri{{\mathscr I^+}} 
\usepackage{graphicx,psfrag}
\usepackage{mathrsfs}

\begin{document}


\title{Asymptotically null slices in numerical relativity:
  mathematical analysis and spherical wave equation tests}

\author{Gioel Calabrese}

\author{Carsten Gundlach}

\author{David Hilditch}

\affiliation{School of Mathematics, University of Southampton,
Southampton, SO17 1BJ, UK}

\date{December 2005, revised April 2006}

\pacs{04.25.Dm, 04.30.Db}


\begin{abstract}

We investigate the use of asymptotically null slices combined with
stretching or compactification of the radial coordinate for the
numerical simulation of asymptotically flat spacetimes. We consider a
1-parameter family of coordinates characterised by the asymptotic
relation $r\sim R^{1-n}$ between the physical radius $R$ and
coordinate radius $r$, and the asymptotic relation $K\sim R^{n/2-1}$
for the extrinsic curvature of the slices. These slices are
asymptotically null in the sense that their Lorentz factor relative to
stationary observers diverges as $\Gamma\sim R^{n/2}$. While $1<n\le
2$ slices intersect $\scri$, $0< n\le 1$ slices end at $i^0$. We carry
out numerical tests with the spherical wave equation on Minkowski and
Schwarzschild spacetime. Simulations using our coordinates with
$0<n\le 2$ achieve higher accuracy at lower computational cost in
following outgoing waves to very large radius than using standard
$n=0$ slices without compactification. Power-law tails in
Schwarzschild are also correctly represented. 

\end{abstract}

\maketitle


\section{Introduction}
\label{Introduction}


Numerical simulations of astrophysical events in general relativity
are carried out on a numerical domain that comprises a central strong
field zone surrounded by a much larger outer zone which contains no
matter, where the fields are weak, and in which gravitational waves
are mostly outgoing.

One reason for making the numerical domain large is
mathematical. Boundary conditions (BCs) for the continuum
initial-boundary value problem which are both compatible with the
Einstein constraints and make the problem well-posed have been
suggested only recently \cite{FriedrichNagy} { and are still untested
numerically. BCs which are consistent with the constraints (but not
proven to be well-posed) \cite{CalabreseLehnerTiglio} have been used
successfully in 3D testbed simulations
\cite{Caltech,SarbachTiglio}. Large simulations of astrophysically
interesting scenarios have so far been using BCs at finite radius
(with the exception of \cite{Pretorius}, see below) that are
incompatible with the constraints, so that { finite constraint
violations propagate in from the boundary even in the limit of
infinite resolution.}

A second reason is physical. The central object should be
isolated with no incoming gravitational radiation. This cannot be
achieved with purely local BCs. For the same reason gravitational
radiation cannot be read off reliably close to the source.

{ The physical problem can be reduced by pushing the outer
boundary to larger radius, and the mathematical problem will
be alleviated at least by making the domain of dependence of the
initial data larger. There is empirical evidence that the dominant
error in astrophysics-type GR simulations is from the boundary, and
that it can be alleviated for example by using nested boxes mesh
refinement to push the outer boundary further out
\cite{MayaFMR,collapseFMR,Miller}.}

A radical approach to both the mathematical and the physical problems
is to represent a complete asymptotically flat spacetime on the
numerical domain. The no ingoing radiation condition can then be
imposed exactly at past null infinity ${\mathscr I^-}$, and the
outgoing radiation can be read off exactly at future null infinity
$\scri$. Regularising the field equations at infinity requires
embedding the Einstein equations into a much larger system called the
conformal field equations \cite{Frauendiener}, but this system has not
yet been used for astrophysical simulations. Alternatively, the
Einstein equations can be regularised at $\scri$ using retarded time
as a null coordinate \cite{Winicour}, but this approach cannot be used
in the central strong field region because the null cones form
caustics. 

{ Finally, space can be compactified on standard
spacelike slices going out to $i^0$. If the initial data are
stationary outside a central region one can then get away with a
pseudo-regularisation of the field equations in their discretised form
at the compactification boundary \cite{Honda,Garfinkle,Pretorius}. In
the continuum equations no gravitational waves reach $i^0$ and no
evolution takes place there. In the discretised equations spurious
radiation must be suppressed by artificial dissipation (or the
implicit dissipation of the discretisation) before reaching the
compactification boundary.}

One key ingredient of the conformal field equations approach is the
use of asymptotically null time slices which terminate at $\scri$,
together with a compactification of the radial coordinate. It has been
suggested to use such coordinates in numerical relativity, but to
introduce an artificial outer boundary at some very large radius and
to use the standard form of the Einstein equations
\cite{Kansagra,Misner}. Fig.~\ref{fig:domain} illustrates this
approach. Ingoing (backscattered) waves at large radius are
increasingly lost { in numerical simulations in this approach because
they are not resolved on any finite resolution grid. It is sometimes
not appreciated that all approaches that put $\scri$ at a finite
coordinate value, in particular the conformal field equations and the
Cauchy-characteristic matching approaches, suffer from the same
shortcoming.} { For example, in Minkowski spacetime with metric
$ds^2=-dUdV + R^2d\Omega^2$ the conformal approach would introduce a
compactification $U=\tan u$, $V=\tan v$ on a grid 
with constant
resolution $\Delta u$ and $\Delta v$. This implies $\Delta V=\sec^2\Delta
v\simeq V^{-2}\Delta v$ (as $V\to\infty$), so that any ingoing waves
with period $f$ are no longer resolved for $V\gtrsim
(f\Delta v)^{-1/2}$. A similar argument holds for the characteristic
approach on outgoing null cones.}

\begin{figure}
\psfrag{a}{$R\lesssim 10M$}
\psfrag{b}{$t=1000M$}
\psfrag{c}{$R\sim 2000M$}
\psfrag{d}{$R\sim 2000M-t$}
\psfrag{e}{$R\sim 1000M$}
\psfrag{f}{$t=0$}
\includegraphics[height=8cm]{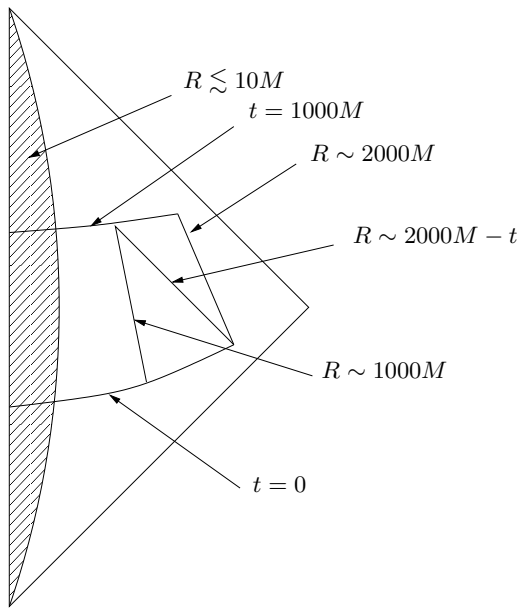} 
\caption{ 
\label{fig:domain}
A schematic spacetime diagram to illustrate that if the artificial outer
boundary is pushed out to sufficiently large radius (here $R=2000M$)
outgoing gravitational radiation can be read off at very large
radius (here $R=1000M$) for a long time (here $t=1000M$) without being
contaminated by unphysical BCs at the artificial outer
boundary (assuming that no information travels faster than light).}
\end{figure}

In this paper, we shall restrict ourselves to the toy model of the
wave equation on Minkowski or Reissner-Nordstr\"om spacetime. In
Sect.~\ref{section:basics} we investigate in some detail the
relationship between radial stretching or compactification, bending up
the slices, the coordinate speeds of light, and the global structure
of the slices. We propose a one-parameter family of asymptotically
null slices comprising flat slices at one extreme and hyperboloidal
slices at the other.

In Sect.~\ref{section:tests} we carry out numerical evolutions of an
outgoing wave packet in the spherical wave equation in ADM-like form
on the Minkowski and Schwarzschild spacetimes. We compare the accuracy
and the numerical cost of evolving on different asymptotically null
slices with the standard slicing, for realistic values of the
numerical parameters. In the Schwarzschild simulations we also look
for power-law tails and find that they are correctly resolved. 
Sect.~\ref{section:conclusions} contains our conclusions.


\section{Mathematical analysis}
\label{section:basics}


\subsection{Spherically symmetric slices in Minkowski spacetime}


The metric of Minkowski spacetime in standard spherical polar
coordinates is
\begin{equation} 
\textrm{d}s^2 =-\textrm{d}T^2+\textrm{d}R^2+R^2{d}\Omega^2,
\end{equation}
where $d\Omega^2\equiv d\theta^2+\sin^2\theta\,d\varphi^2$.

{ Consider the flat space wave equation as a toy model for the
linearised Einstein equations.} An outgoing wave $\phi$ emanating from
a central region can be approximated as
\begin{equation}
\phi(T,R,\theta,\varphi)\simeq \sum_{l=0}^{l_{\rm max}}\sum_{m=-l}^l
\phi_{lm}(U) R^{-1} Y_{lm}(\theta,\varphi) 
\end{equation}
for some small finite $l_{\rm max}$, where the $Y_{lm}$ are spherical
harmonics and $U=T-R$ is retarded time. To resolve this, constant
spatial resolution $\Delta x$, $\Delta y$, $\Delta z$ is not required,
but only constant angular resolution $\Delta\theta$,
$\Delta\varphi$. Similarly, constant resolution in $R$ and $T$
separately is not required, but only constant resolution in $U$. If
the slices of constant coordinate time $t$ ``bend up'' such that they
approximate slices of constant $U$ asymptotically as $R\to\infty$,
then the radial grid spacing $\Delta R$ can be allowed to diverge in
this limit.

To study this quantitatively, we make the change of coordinates
\begin{eqnarray}
\label{slice}
t &=& T-F(R), \\
\label{radtrans}
R &=& R(r),
\end{eqnarray}
where $F(R)$ controls the slicing, and $R(r)$ the spatial coordinates
on the slices. The new slicing is static, that is, Lie-dragged by the
Killing vector $\partial/\partial T$.

The metric in 3+1 in form in the new coordinates is
\begin{equation}
\textrm{d}s^2 =-\alpha^2\textrm{d}t^2+\gamma_{rr}(\textrm{d}r
+\beta^r\textrm{d}t)^2+\gamma_{\theta\theta}\textrm{d}\Omega^2,
\end{equation}
with the lapse, shift, 3-metric and extrinsic curvature given by 
\begin{eqnarray}
\alpha &=& \frac{1}{(1-F'^2)^{\frac{1}{2}}}, \\ 
\beta^r &=& -\frac{F'}{R'(1-F'^2)}, \\
\gamma_{rr} &=& R'^2(1-F'^2), \\ 
\gamma_{\theta\theta} &=& R^2, \\ 
K_{rr} &=& -\frac{F''R'^2}{(1-F'^2)^\frac{1}{2}},  \\
K_{\theta\theta} &=& -\frac{F'R}{(1-F'^2)^{\frac{1}{2}}}. 
\end{eqnarray}
Note that $\alpha\ge 1$. { Here and in the following $F'\equiv
  dF/dR$ and $R'\equiv dR/dr$.}
The coordinate light speeds in the radial and tangential directions
are
\begin{eqnarray}
c_{\pm} &\equiv&
\left.\frac{\textrm{d}r}{\textrm{d}t}\right|_{\theta,\varphi}
=\pm {\alpha\over \sqrt{\gamma_{rr}}}-\beta^r
= \pm\frac{1}{R'(1\mp F')}, \\ 
c_{\theta} &\equiv &
\left.\frac{\textrm{d}\theta}{\textrm{d}t}\right|_{r,\varphi} 
= \sqrt{-g_{tt}\over \gamma_{\theta\theta}}
= \frac{1}{R}.
\end{eqnarray}
We can also express the radial coordinate speeds of light
$dr/dt=c_\pm$ directly in terms of $dR/dT=C_\pm$ in the form
\begin{equation}
\label{cC}
c_\pm^{-1}=R'(C_\pm^{-1}-F'),
\end{equation}
and this relation holds for any spherically symmetric metric.


\subsection{Stability conditions}


We consider a numerical stencil centred at the grid point $x^i_0$ at
time $t_0$, and for simplicity of presentation we shift the origin of
the coordinates so that $x^i_0=0$ and $t_0=0$. We model the stencil as
the square box $-\Delta x^i\le x^i \le\Delta x^i$ for $i=1,2,3$. The
time step is $\Delta t$, and for simplicity of notation we assume a
two-level scheme and a three-point stencil in each spatial direction.

The intersection of the past light cone of the grid point at $x^i=0$
and $t=\Delta t$ with the time slice $t=0$ is the ellipsoid
\begin{equation}
\gamma_{ij}(x^i+\beta^i \Delta t)(x^j+\beta^j \Delta t)=\alpha^2
\Delta t^2.
\end{equation}
Note that the origin of this ellipsoid is shifted from $x^i=0$ if
$\beta^i\ne 0$, and its principal axes are not aligned with the
coordinate axes if $\gamma_{ij}$ is not diagonal.

The light ellipsoid fits into the stencil box (the
Courant-Friedrichs-Levy (CFL) condition) if and only if in each of the
three coordinate directions $i$ the maximal value of $x^i$ on the
ellipsoid is less than $\Delta x^i$ and the minimal value is greater
than $-\Delta x^i$. The solution of the resulting extremisation
problem gives
\begin{equation}
{\Delta x^i \over \Delta t}\ge \alpha\sqrt{\gamma^{ii}}+|\beta^i|
\end{equation}
for $i=1,2,3$ (no summation over $i$). This result is valid also for
non-diagonal $\gamma_{ij}$. 

Restricted to spherical symmetry, these conditions become
\begin{eqnarray}
{\Delta r\over \Delta t}
&\ge& {\alpha\over\sqrt{\gamma_{rr}}}+|\beta^r|, \\
{\Delta \theta\over \Delta t}
&\ge& {\alpha\over R}, \\
{\Delta \varphi\over \Delta t}
&\ge& {\alpha\over R|\sin\theta|}.
\end{eqnarray}
{ The factor $1/\sin\theta$ in the third condition reflects the
standard problem with spherical coordinates on the $z$-axis unrelated
to our coordinates and which we ignore here.}

The CFL condition, which is necessary for stability, can be written as
\begin{equation}
|c_+|\le k{\Delta r\over \Delta t} , \quad 
|c_-|\le k {\Delta r\over \Delta t}, \quad
c_T\le k {\Delta \theta\over \Delta t}
\end{equation}
for $k=1$, where
\begin{equation}
c_T\equiv {\alpha\over R}= {1\over R(1-F'^2)^{1/2}}.
\end{equation}
For typical explicit finite differencing methods for linear hyperbolic
equations, a sufficient stability condition is then of the same form
for some constant $k$ which depends on the equation and its
discretisation (allowing for any number of time levels and size of
stencil). This will generalise to any hyperbolic formulation of the
Einstein equations. Note that $c_T\ne c_\theta$ unless $\alpha=1$, and
that $c_T$ and not $c_\theta$ is the relevant quantity for stability.


\subsection{Slicing, stretching, compactification, and lightcones}


The assumption that the slice is spacelike is equivalent to $F'<1$,
and the assumption that it becomes asymptotically null is equivalent
to $F'\to 1$. For simplicity let us consider functions $1-F'$ that
decay as a power of $R$ at large $R$, or
\begin{equation}
\label{fpower}
1-F'(R)\sim R^{-n}
\end{equation}
as $R\to\infty$, where $n>0$. Without loss of generality we assume
that the numerical grid is equally spaced in $r$. Then the CFL
condition in the radial direction requires that
\begin{equation}
c_+\sim {R^n\over R'}
\end{equation}
is bounded above as $R\to\infty$. To preserve accuracy, it should also
be bounded below. (If $c_+\to 0$ as $R\to \infty$, an outgoing wave
pulse would slow down, become increasingly narrow, and be damped by
numerical dissipation.) Therefore
we require 
\begin{equation}
c_+\sim {R^n\over R'}\sim \hbox{const.},
\end{equation}
which on integration gives us
\begin{equation}
r\sim R^{1-n}.
\end{equation}
This means that for $0<n<1$ we have radial stretching
\begin{equation}
R(r)\sim r^{1/(1-n)}, \quad r\to\infty, 
\end{equation}
and for $n>1$ we have radial compactification
\begin{equation}
R(r)\sim (l-r)^{-1/(n-1)}, \quad r\to l_-.
\end{equation}
The special case $n=1$ gives us
\begin{equation}
R(r)\sim e^r, \quad r\to\infty,
\end{equation}
also a case of stretching.

We see that
\begin{equation}
c_- \simeq -{1\over 2R'}\to 0
\end{equation}
as $R\to\infty$ for either stretching or compactification.  This means
that ingoing wave pulses move very slowly (in $r$ and $t$) at large radius,
and that they become wider (in $r$) as they move in. 
Although they are insufficiently
resolved at large radius, they do not pose a stability problem for the
finite differencing scheme. We also see that
\begin{equation}
\label{cTlimit}
c_T\sim R^{n/2-1}
\end{equation}
as $R\to\infty$, and so with constant angular resolution the CFL
condition in the tangential directions requires $n\le 2$.

In summary, the ansatz (\ref{fpower}) for the slicing requires $0<n\le
2$ for the coordinate speeds in all directions to be bounded. Keeping
$c_+$ bounded below as well as above means that $0<n\le 1$ requires
radial stretching and $1<n\le 2$ requires radial compactification.

As a matter of convention, we always set $F(0)=0$, $R(0)=0$ and
$R'(0)=1$, so that $c_+(0)=1$. For regularity we require $R(r)$ to be
odd and $F(R)$ to be even.  We introduce a length scale $L$ so that
\begin{equation}
\label{RprimeLn}
R'\simeq \left({R\over L}\right)^n
\end{equation}
as $R\to\infty$. We demand that $c_+(\infty)=1$, which implies
\begin{equation}\label{slicerequires}
F'\simeq 1-\left({R\over L}\right)^{-n}.
\end{equation}
Simple closed form expressions for $F(R)$ and $R(r)$ with the required
asymptotic behaviour are given in App.~\ref{appendix:closedform}. A
typical set of coordinate speeds is plotted in
Fig.~\ref{fig:KSweakslicespeeds}. 

In summary, compactifying the radius without bending up the slices
would mean losing the outgoing radiation.
Bending up the slices without radial compactification would increase
the radial outgoing characteristic speeds without limit and so violate
the CFL condition in the radial direction. Bending up the slices
without keeping the angular resolution constant at large radius would
violate the CFL condition in the angular directions.


\subsection{Global structure}


The exponent $n$ can be given a geometric interpretation as
follows. There are two preferred families of observers: those normal
to the slicing and those along the Killing field $\p/\p T=\p/\p t$. The speed
of the former relative to the latter is 
\begin{equation}
v=-{\beta^r\sqrt{\gamma_{rr}}\over\alpha},
\end{equation}
and the corresponding Lorentz factor is
\begin{equation}
\Gamma\equiv {1\over \sqrt{1-v^2}}\simeq {1\over \sqrt{2}}\left({R\over
  L}\right)^{n/2}
\end{equation}
as $R\to\infty$. 

We see that our $n>0$ slices are asymptotically null in the precise
sense that their Lorentz factor diverges as $R\to\infty$, but
surprisingly this does not imply that they reach $\scri$. Consider the
null coordinates $V\equiv T+R=t+F(R)+R$ and $U\equiv T-R=t+F(R)-R$ on
Minkowski spacetime. { Noting that $F(R)\simeq
R$ as $R\to\infty$, we find that $V\simeq 2R\to\infty$ as $R\to\infty$
on a slice of constant $t$. For $U$ on a slice of constant $t$, we
find that
\begin{equation}
\left.{dU\over dR}\right|_{t={\rm
    const.}}=F'(R)-1\simeq -\left({R\over L}\right)^{-n}
\end{equation}
Integrating this relation over $R$, we find that for $n>1$, $U$
approaches a finite value as $R\to\infty$ (or equivalently, as we have
seen, $V\to\infty$)} on a slice of constant $t$, and so the slice
reaches a point on $\scri$. For $n\le 1$, $U\to -\infty$ as
$V\to\infty$ on a slice of constant $t$, and so the slice reaches
$i^0$.  If our slices are plotted on the standard conformal diagram,
they approach $i^0$ horizontally for $n=0$, but tangentially to
$\scri$ for $0<n\le 1$. For $1<n\le 2$ they intersect $\scri$.

For comparison we determine the asymptotic behaviour of the standard
conformal compactification of Minkowski \cite{HawkingEllis}. With $U$
and $V$ as above, and $u\equiv (t-r)/2$ and $v\equiv (t+r)/2$, this is
given by $U=\tan u$ and $V=\tan v$. (For consistency with the rest of
this paper $u$ and $v$ have finite range and $U$ and $V$ infinite
range). Clearly this is is not of the restricted form we have
considered here and which gives a static slicing. But we can show that
as $R\to\infty$ on a slice of constant $t$,
\begin{equation}
\left.{dT\over dR}\right|_{t={\rm
    const.}}\simeq 1-\left({R\over L(t)}\right)^{-2}
\end{equation}
which we can interpret as $n=2$ but where $L(t)$ now depends
explicitly on the slice. In this compactification $c_\pm=\pm 1$
exactly, and so both ingoing and outgoing waves are represented
accurately. The lines of constant $(r,\theta,\varphi)$ are no longer
Killing trajectories, and so we cannot implement this gauge choice
with ``symmetry-seeking'' coordinate conditions \cite{symmcoord},
which include almost all popular coordinate conditions.


\subsection{Reissner-Nordstr\"om spacetime}
\label{section:mass}


To obtain asymptotically null slices in Reissner-Nordstr\"om
spacetime, we start from its metric in (for example) Kerr-Schild coordinates
$(T,R)$,
\begin{eqnarray}
\label{KSasym}
ds^2 &=& -(1-f)\,dT^2+2f\,dT\,dR+(1+f)\,dR^2\br 
+ R^2\,d\Omega^2 \quad \hbox{where } f(R)\equiv {2m\over R}-{q^2\over R^2},
\end{eqnarray}
and make a coordinate transformation of the form
(\ref{slice},\ref{radtrans}) to new coordinates $(t,r)$. We choose
$R(r)$ to have the asymptotic behaviour (\ref{RprimeLn}) with $0< n\le
2$. From (\ref{cC}) we find that in order to obtain $c_+(\infty)=1$,
the asymptotic behaviour of $F(R)$ as $R\to 0$ must now be
\begin{equation}
\label{FprimeSchwarz}
F'= 1+{4m\over R}+{8m^2-2q^2\over R^2}-\left({R\over
  L}\right)^{-n}+o(R^{-n}),
\end{equation}
for $0<n\le 2$. For $n<2$ the $R^{-2}$ term in this expression can be
omitted, and for $n<1$ the $R^{-1}$ term can also be omitted, as they
give only $o(1)$ contributions to $c_+$. Conversely, we see that for
$n\ge 1$ the slicing must be corrected for the total mass of the
isolated central object. In a dynamic 3D spacetime this mass would
presumably be the ADM mass for $n=1$ and the Bondi mass for
$n>1$. Although our calculations are limited to spherical symmetry,
the analogy between charge and angular momentum suggests that with
$n=2$ in 3D the slicing would have to be corrected for the Bondi
angular momentum as well. 

Finally we note that (\ref{cTlimit}) holds also in
Reissner-Nordstr\"om spacetime. More precisely,
\begin{equation}
c_T\simeq {1\over \sqrt{2}R}\left({R\over L}\right)^{n/2}
\end{equation}
for $0<n<2$ with any $m$ and $q$, and 
\begin{equation}
c_T\simeq {1\over \sqrt{2} \tilde L}, \quad \tilde L^2
\equiv L^2+24m^2-4q^2
\end{equation}
for $n=2$ as $R\to\infty$.

\begin{figure}
\begin{center}
\includegraphics[width=0.45\textwidth]{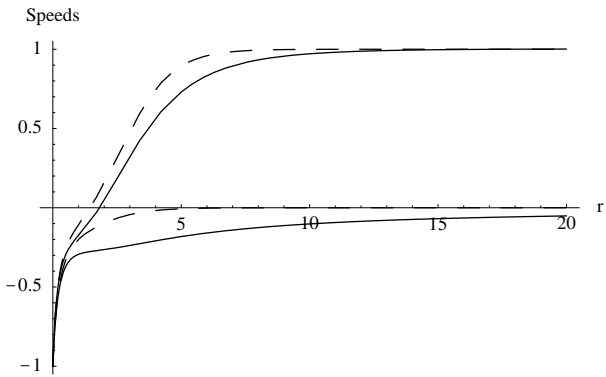}
\caption{\label{fig:KSweakslicespeeds}\footnotesize{Coordinate speeds
$c_+$ and $c_- $ in modified Kerr-Schild coordinates, made
asymptotically null with $n=\frac{1}{2}$ (solid lines) and $n=1$
(dashed lines). $L=1$ in both cases.}}
\end{center}
\end{figure}


\subsection{Characterisation of $n$ in terms of 3+1 variables}
\label{section:3+1}


We have already characterised the slicing geometrically in terms of
its Lorentz factor. Another geometric characterisation is in terms of
the metric and intrinsic curvature of the slices. Applying the
coordinate transformation (\ref{slice},\ref{radtrans}) with the
asymptotic behaviour (\ref{RprimeLn},\ref{FprimeSchwarz}) to the
metric (\ref{KSasym}), we find that for Minkowski with $0<n\le 2$ and
Reissner-Nordstr\"om with $0<n<2$,
\begin{eqnarray}
\alpha &\simeq& {1\over \sqrt{2}}\left({R\over L}\right)^{n\over 2}, \qquad
\beta^r \simeq -{1\over 2}, \\
\gamma_{rr} &\simeq & 2 \left({R\over L}\right)^n, \qquad
\gamma_{\theta\theta} \equiv R^2, \\
\label{Krr}
K_{rr} &\simeq& -{n\over \sqrt{2}L} \left({R\over L}\right)^{{3n\over
    2}-1}, \\ 
K_{\theta\theta} &\simeq& -{L\over \sqrt{2}}\left({R\over
  L}\right)^{{n\over 2}+1}, \\
K &\simeq & -{n\over 2\sqrt{2}L} \left({R\over L}\right)^{{n\over 2}-1},
\end{eqnarray}
in the limit $R\to\infty$. (For $n=2$ with $m,q\ne 0$, the leading
order coefficients change from the ones given here).  $R$ is defined
geometrically as the area radius. To leading order in $R$, there is no
effect from the mass $m$. As $R\to\infty$,
$K={K^r}_r+2{K^\theta}_\theta$ is dominated by ${K^r}_r$ for all
$n$. For $n=0$, $K$ vanishes identically, for $0<n<2$ it goes to zero
as $R\to\infty$, while for $n=2$ it approaches a constant.


\section{Numerical Experiments}
\label{section:tests}


\subsection{Numerical method}


We solve the wave equation restricted to spherical symmetry on
Minkowski and Schwarzschild spacetime. Our initial data is an outgoing
Gaussian pulse near the origin (but outside the black hole). On Minkowski
spacetime the problem has an overall scale-invariance, which we fix by
setting the width of the pulse to $\Delta R\sim 1$. In the Schwarzschild 
simulations, we also set $m=1$.

In astrophysical simulations we would need to maintain a fairly
constant resolution in a finite inner region where the dynamics take
place. As a testbed for 3-dimensional simulations, we therefore use a
resolution of $\Delta r=0.1$, which, with our convention $R'(0)=1$,
corresponds to $\Delta R\simeq 0.1$ in the central region, or $\sim
10$ grid points over the width of the pulse. We therefore choose $L$
or $l$ so that $R'(r)\simeq 2$ at $R=10$, which means that resolution
at that radius is still half of that at the centre.

We solve the wave equation { in the continuum form
(\ref{wave1},\ref{wave2})}. We use 2nd and 4th-order accurate finite
differencing schemes proved to be stable in \cite{boundshifted}, with
a Courant factor of $\Delta t/\Delta r=0.4$. A regular origin is dealt
with by imposing regularity, and taking appropriate limits in the
right hand sides. For black hole excision we use outflow BCs.

We estimate the error and check that the results converge to the
expected order by comparing the three resolutions $\Delta r=0.1$,
$0.05$ and $0.025$. An example is Fig.~\ref{rescaledn1}, where
we plot the local error $e(r,t_*)$, where $t_*\equiv t(R=R_*,
T=R_*)$ is the value of $t$ when the maximum of the
Gaussian pulse is at physical radius $R=R_*$ in a given slicing.


\subsection{Outer boundary treatment}


If the outer boundary is at $R_{\rm max}=\infty$, we impose BCs from the exact
solution. For the 2nd-order accurate code, this simply means imposing
$\phi=\Pi=0$ at the outermost grid point $i=N$ where $R=\infty$. In the
4th-order accurate code on Minkowski we also impose the exact solution
at $i=N-1$. On Schwarzschild, where no exact solution exists, we
copy $R\phi$ and $R\Pi$ from $i=N-2$ to $i=N-1$. (This is only
first-order accurate, but in practice does not affect results away
from the boundary.)

If the outer boundary is at finite physical radius $R_{\rm max}$, we
impose maximally dissipative BCs (MDBCs) on $\phi$ and $\Pi$, with
lower-order terms representing the $1/R$ falloff. This continuum BC
perfectly represents outgoing waves in spherical symmetry on flat
spacetime, but in curved spacetime and/or for non-spherical waves
gives rise to an unavoidable small continuum reflection. The finite
differencing implementation of this BC also gives rise to numerical
reflections which converge away with increasing resolution.

With $n>0$ the outer BC is ``almost'' an outflow boundary. A second
possible BC is therefore to increase the radial shift
by a small amount, leaving the other metric coefficients unchanged, to
turn the outer boundary into a genuine outflow boundary. The
background spacetime is then no longer a vacuum solution of Einstein's
equations but the error is small. This is a simple variant of Misner's
idea \cite{Misner} of introducing a small cosmological constant to
create a cosmological horizon that represents a null outer boundary at
constant $R$.

As a third BC, we have also tested an outer buffer zone in which $c_+$ goes
smoothly to zero. This can be achieved by reducing $n$ of the slicing
and/or increasing $n$ of the radial coordinate transformation. In
either case the effect is that outgoing waves become ``blue-shifted''
with respect to the grid spacing and are then suppressed by artificial
dissipation. We consider the buffer not as part of the physical
spacetime but rather as a BC. For convenience of implementation, with
$n\ge1$ in the physical region we reduce $n$ for the slicing in the
buffer, and with $n<1$ in the physical region we increase $n$ of the
radial coordinate transformation (to a value $>1$) in the buffer. This
means that for all values of $n$ in the physical region the buffer
goes to $\infty$.


\subsection{Spherical wave equation on Minkowski}


We evolve an outgoing scalar field pulse with different slicings with
the aim of quantifying the error and establishing what fraction of
grid points can be saved by using asymptotically null coordinate
slices, and what effect they have on the error as the wave goes out to
very large $R$. We present results for
$n=1/2,1,3/2,2$. The results for $n\le 1$ have been
obtained with MDBCs at $R\gtrsim 1000$, and the results for $n>1$ by
imposing $\phi=\Pi=0$ at $\scri$ and, in 4th-order accuracy, the exact
solution at the grid point just inside the boundary.

\begin{table*}
\begin{tabular}{|c|cc|cc|ccc|ccc|}
\hline
 {\small  Slice                           }
&\multicolumn{2}{|c|}{{\small Grid points to cover}} 
&\multicolumn{2}{|c|}{{\small Time steps to reach }} 
&\multicolumn{3}{|c|}{{\small $|e(\cdot,t_*)|$ with 2nd order code at}} 
&\multicolumn{3}{|c|}{{\small $|e(\cdot,t_*)|$ with 4th order code at}} 
\\
          ${}$ 
& {\small  $R_{\rm max}=1000$             } 
& {\small  $R_{\rm max}=\infty$            }
& {\small  $R_*=1000$                     }
& {\small  $R_*=\infty$                   }
& {\small  $R_*=10$                       }
& {\small  $R_*=100$                      }
& {\small  $R_*=1000$                     }
& {\small  $R_*=10$                       }
& {\small  $R_*=100$                      }
& {\small  $R_*=1000$                     }
\\
\hline
  $n=0$ 
& $10000$
& $\infty$ 
& $25000$
& $\infty$
& $1.8\cdot10^{-3}$
& $1.7\cdot10^{-3}$
& $5.0\cdot10^{-4}$
& $2.0\cdot10^{-5}$
& $1.9\cdot10^{-5}$
& $1.9\cdot10^{-5}$
\\
  $n=\frac{1}{2}$
& $893$
& $\infty$
& $2300$
& $\infty$
& $1.5\cdot10^{-3}$
& $7.5\cdot10^{-4}$
& $2.5\cdot10^{-4}$
& $1.7\cdot10^{-5}$
& $1.2\cdot10^{-5}$
& $4.5\cdot10^{-6}$
\\
$n=1$
& $338$
& $\infty$
& $1000$
& $\infty$
& $2.3\cdot10^{-3}$
& $8.0\cdot10^{-4}$
& $1.5\cdot10^{-4}$
& $3.2\cdot10^{-5}$
& $1.3\cdot10^{-5}$
& $2.0\cdot10^{-6}$
\\
  $n=\frac{3}{2}$ 
& $207$
& $224$
& $624$
& $667$
& $2.5\cdot10^{-3}$
& $1.1\cdot10^{-3}$
& $2.5\cdot10^{-4}$
& $2.3\cdot10^{-5}$
& $2.4\cdot10^{-5}$
& $4.8\cdot10^{-6}$
\\
  $n=2$
& $165$
& $167$
& $390$
& $418$
& $1.5\cdot10^{-3}$
& $6.0\cdot10^{-4}$
& $5.0\cdot10^{-4}$
& $1.8\cdot10^{-5}$  
& $1.0\cdot10^{-5}$
& $4.0\cdot10^{-6}$
\\
\hline
\end{tabular}
\caption{Computational costs and errors in evolving a spherical wave
on Minkowski spacetime for different slicings. As the outgoing wave
has amplitude $1/R$, a suitable measure of the {\em relative} error
would be $e/R$.}\label{table:gridpoints}
\end{table*}

We evolve up to the time when the outgoing wave reaches the outer
boundary. Until then solutions for all $n$ converge with the expected
order (2nd or 4th), and so we have an accurate estimate of the
numerical error. The number of grid points required to reach a given
physical radius $R_{\rm max}$ and the norm of the finite differencing
error when the wave has gone out to three selected values $R_*$ of $R$ are
summarised in Table~\ref{table:gridpoints}. Fig.~\ref{rescaledn1} is a
typical example of the error plots.

The larger the value of $n$, the smaller the number of grid points
required. The error when the wave reaches $R=10$ is similar for all
values of $n$ as one would expect from the fact that all grids are
similar in the central region $R\lesssim 10$, but by the time the wave
reaches $R=1000$ all $n>0$ slicings do better than $n=0$, with $n=1$
an order of magnitude more accurate than $n=0$. We believe that this
could be explained by the much smaller number of time steps required
for the wave to reach $R=1000$, for example 338 for $n=1$ compared to
10000 for $n=0$. This would tend to reduce phase error, and in
dissipative schemes also amplitude error.

\begin{figure}
\begin{center}
\includegraphics[width=0.45\textwidth]{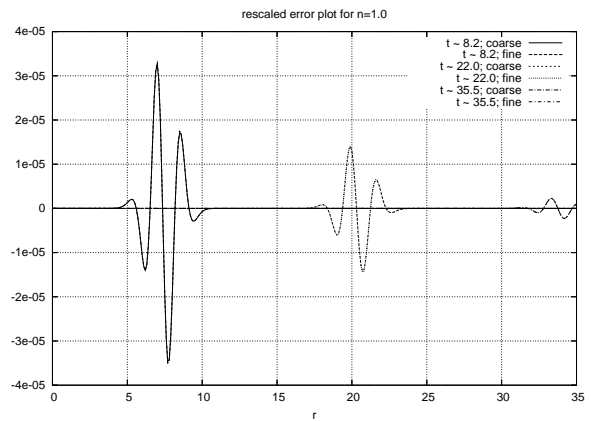}
\caption{\label{rescaledn1}\footnotesize{The Error in $\phi$ with
$n=1$ and $L=5.77$, at $R_*=10$, $100$ and $1000$.  $e(t,r_*)$ at
$\Delta r=0.1$ and $16e(r,t_*)$ at $\Delta r=0.05$ are both plotted,
but are indistinguishable here, indicating 4th-order convergence. The
absolute error $e$ decreases with $R_*$, but the relative error $e/R$
increases.}}
\end{center}
\end{figure}


\subsection{Spherical wave equation on Schwarzschild}


Fields in a curved asymptotically flat spacetime generically decay
with power law tails because of backscatter \cite{Price}. In order to
see how well our coordinates can cope with such small physical effects
at large physical radii, we evolve the spherical wave equation on the
Schwarzschild background, on the horizon-penetrating asymptotically null
slices derived above. A priori it is not clear if tails will be
represented correctly or not: on the one hand they are caused by
backscattered {\em ingoing} waves, on the other hand their frequency in
advanced time is low.

The numerical algorithm is the same as for our Minkowski experiments,
except for the boundaries. The regular origin is replaced by an
outflow (excision) boundary inside the black hole. We do not see large
errors or any sign of instability associated with the excision
boundary. We use the 4th-order
accurate scheme with the artificial dissipation parameter
$\sigma=0.007$. The results depend only weakly on $\sigma$ { (including
$\sigma=0$)}. 

At the outer boundary we use three types of BC. The first
corresponds to what we did in Minkowski spacetime, that is, we impose
MDBCs at finite $R_{\rm max}$ or ``exact'' BCs at $\infty$. This gives
good tail results if we put the outer boundary far enough out. As a
second type of boundary we have increased the shift near the boundary
slightly to obtain an outflow BC. As a third type of BC we have added
a $20\%$ buffer in which $c_+$ goes to zero and $R$ reaches infinity
(even if $n\le 1$ in the physical region). The case $n=0$ with such a
buffer is similar to compactification at $i^0$ used in
\cite{MayaFMR,collapseFMR,Miller}.

As we do not have an exact solution, we have read off initial data
$\phi$ and $\Pi$ on a given initial slice from what would be an exact
solution on Minkowski where the Minkowski coordinates $T$ and $R$ are
identified with Kerr-Schild coordinates. This means that we have
slightly different initial data for different slicings. We have
therefore used a high resolution simulation on each slicing as a
reference solution to decide when power-law tails are lost. We see
that at $R=500$ the power is not exactly 3 as expected for constant
$R$, but varies between 3 and 2 (the power expected at $\scri$).

\begin{table*}
\begin{tabular}{|c|c|c|cc|cc|cc|}
\hline
 {\small Slice                                     }
&{\small $R_{\rm  max}$                            }
&{\small Grid points                               }
&\multicolumn{2}{|c|}{{\small  $t_l$ with MDBCs         }} 
&\multicolumn{2}{|c|}{{\small  $t_l$ with outflow BCs          }}
&\multicolumn{2}{|c|}{{\small  $t_l$ with buffer BCs       }}
\\
  $$
&
& {\small   $$                                                 }
&\multicolumn{2}{|c|}{ {\small or exact BCs at $\infty$        }}
&\multicolumn{2}{|c|}{ {\small                                }}
&\multicolumn{2}{|c|}{ {\small                   }}
\\
&&
& {\small   $R=10$                                               }
& {\small   $R=500$                                              }
& {\small   $R=10$                                               }
& {\small   $R=500$                                              }
& {\small   $R=10$                                               }
& {\small   $R=500$                                              }
\\
\hline
  $n=0$
& $1000$
& $10086$
& $2020$
& $1530$
& 
& 
& $>2100$ 
& $>2100$
\\
\hline
  $n=\frac{1}{2}$
& $1000$
& $884$
& $\sim1700$
& $\sim900$
& $\sim1800$
& $\sim900$
& $>2100$
& $>2100$
\\

& $10^6$
& $28271$
& $>10^4$
& $>10^4$
& $>2800$ 
& $>2800$ 
& 
& 
\\
\hline
  $n=1$
& $1000$
& $324$
& $\sim900$
& $\sim500$
& $\sim600$
& $\sim200$
& $\sim900$
& $\sim500$
\\

& $10^6$
& $722$
& $\bf >10^4$
& $>10^4$
& $>2400$
& $>2400$
& 
& 
\\
\hline
$n=\frac{3}{2}$
& $1000$
& $194$
& $\sim200$
& *
& $\sim200$
& *
& $\sim200$
& *
\\ 

& $\infty$
& $210$
& $\sim300$
& $\sim150$
& 
& 
& 
& 
\\
\hline
$n=2$
& $1000$
& $152$
& *
& *
& $\sim100$
& *
& *
& *
\\

& $\infty$
& $153$
& $\sim300$
& $\sim200$
& 
& 
& 
& 
\\
\hline
\end{tabular}
\caption{Time $t_l$ when power-law tails are lost for different
slicings, different BCs, two outer radii, and two observer
locations. Note that $T=t+\rm const$ for these observers at fixed $R$,
where the constant depends on $R$ and the slicing. A * means that
power-law tails were not seen. A blank means that this combination of
slice and BC is not sensible. $R_{\rm max}$ is the outer radius of the
physical region (defined arbitrarily as $c_+>1/2$), not of the
buffer. With buffer BCs the number of grid points increases by about
20\% from the value stated in the third column.
\label{table:dod}}
\end{table*}

\begin{figure}
\begin{center}
\includegraphics[width=0.45\textwidth]{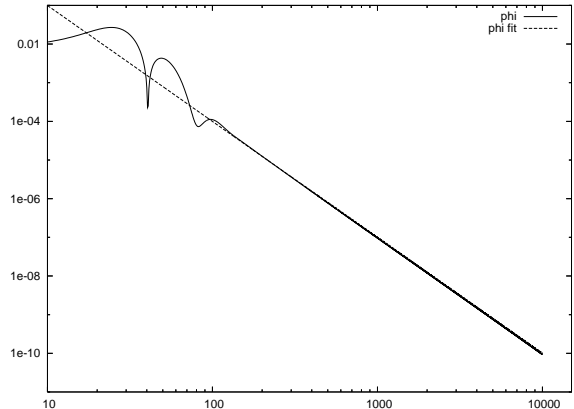}
\caption{\label{KSann1TailsR10}\footnotesize{Log-log plot of $\phi$
versus $t$ at $R=10$ on Schwarzschild, using $n=1$ slices, with an
outer boundary $R_{\rm max}\sim10^6$. The straight line is $t^{-3}$ with
the amplitude but not the power adjusted to the data. This is the
expected power-law tail. The corresponding entry in
Table~\ref{table:dod} is in bold face.}}
\end{center}
\end{figure}

The errors and computational costs on the domain of dependence of the
initial data are similar to those given in
Table~\ref{table:gridpoints}. Therefore we concentrate on power
law tails at late times. Our results are summarised in
Table~\ref{table:dod}. To check for tails, we plot $\phi(t)$ for
``observers'' at $R=10$ and $R=500$. { An example is
Fig.~\ref{KSann1TailsR10}.} It is worth keeping in mind at
what time the boundary can have a physical influence on the
observers. For $n=0$, an observer near the centre leaves the domain of
dependence of the initial data at $t\simeq R_{\rm max}$, while any
continuum reflection of the initial outgoing pulse reaches the observer at
$t\simeq 2R_{\rm max}$. With $n>0$, one can roughly think of the
slices as null, and of outgoing waves as going out almost
instantaneously, so that both events happen close together at $t\simeq
2R_{\rm max}$. We find that for $n=0$, the tail is lost at {\em exactly} the
time when the continuum reflection of the outgoing wave reaches the
observer (at $R=10$ and $500$ respectively). For $n>0$, the tail can
be lost both earlier (presumably because numerical boundary error
travels faster than light) or later (presumably because at very large
boundary radius the continuum reflection is small and the numerical
reflection is also small). 

Our best results are obtained for $n=1$ with MDBCs imposed at $R_{\rm
max}\simeq 10^6$, which requires only 722 grid points for a central
resolution of $\Delta R=0.1$. From the Minkowski results we see that
this choice would also be among the best ones for minimising the error
in the outgoing wave.


\section{Conclusions}
\label{section:conclusions}


The essential motivation for using either the conformal field
equations or null coordinates is their ability to simulate an
asymptotically flat spacetime out to $\scri$ on a finite { domain.
However, this can be done accurately with a finite number of {\em grid
points}} only in situations where ingoing radiation, including
backscatter, can be progressively neglected at large
radius. Conversely, in any physical situation where these approaches
work, one should be able to simulate the spacetime {\em almost} up to
$\scri$ by using asymptotically null slices without regularising the
field equations at infinity \cite{Kansagra,Misner}.

From the requirement that the outgoing coordinate speed of light
should neither go to zero (the code becomes inefficient and too
dissipative) nor to infinity (explicit schemes become unstable, and
implicit schemes are likely to become inaccurate), we have established
a quantitative relationship between the rate at which the slices
become null at large radius and the rate at which the radial
coordinate is compactified.

This rate can be characterised by a parameter $n$ in the range $0\le
n\le 2$, where on the one hand the physical radius (area radius) $R$
and coordinate radius $r$ are asymptotically related by $r\sim
R^{1-n}$, and on the other hand the Lorentz factor of the slices
asymptotically diverges as $\Gamma\sim R^{n/2}$ and the extrinsic
curvature of the slices scales as $K\sim R^{n/2-1}$. 

The extreme value $n=2$ corresponds to the hyperboloidal slices which
have traditionally been used both in drawing the standard conformal
diagram of Minkowski or Schwarzschild spacetime, and in numerical
evolutions using the conformal field equations. For $0<n\le
1$ the range of the coordinate $r$ is infinite (we prefer to speak of
stretching rather than of compactification) and surprisingly the
slices terminate at $i^0$. For $1<n\le 2$ the range of $r$ is finite
and the slices intersect $\scri$ as expected.

Our numerical simulations of the spherical wave equation on Minkowski
and Schwarzschild spacetimes have confirmed our expectations. For a
given (very large) radius of the outer boundary and a given
(realistic) accuracy requirement as the outgoing wave reaches that
boundary, simulations in our $n>0$ coordinates require dramatically
fewer grid points.  To compare two extreme examples, evolving an
outgoing wave with wavelength $\Delta R\sim 1$ out to $R=1000$ with a
relative error of 20\% requires 10000 gridpoints (and 25000 timesteps)
in standard coordinates using 4th-order finite differencing, while
only 165 grid points (and 390 time steps) with our $n=2$ coordinates
give a relative error of only 4\%.

Although $n>0$ coordinates do not represent ingoing waves accurately,
power-law tails of waves on Schwarzschild are correctly represented,
until numerical error from the boundary becomes bigger than the tail
signal. Lower values of $n$ represent the tails correctly for longer
time because numerical error from the boundary propagates in less
rapidly. This suggests that in applications to numerical relativity
all values of $n$ should be considered. From our numerical
experiments, the preferred value of $n$ as a compromise between
minimising grid size, representing outgoing waves, and representing
power-law tails is $n=1$. It also appears that an $n=2$ slicing would
have to be adapted to the angular momentum of the central source,
whereas $n<2$ slicings are only affected by the mass.

In 3D, our method requires approximately constant angular resolution
as $R\to\infty$.  A promising technique for achieving this while
avoiding the axis singularity of spherical polar coordinates is the
use of multiple coordinate patches \cite{CN,CN2,Tiglio,Thornburg}.

In summary, our proposed approach for pushing the BCs to very large
radius requires only a modification of the initial data and the gauge
conditions but uses standard 3+1 field equations. It is simpler than
the conformal field equation method and Cauchy-characteristic matching
methods. Although it cannot reach $\scri$, it can represent outgoing
waves out to arbitrarily large radius on a finite grid. Our
results also suggest certain improvements to existing numerical
relativity codes:

a) A lower value of $n$, perhaps $n=1$, may be better than $n=2$ for
representing physically interesting scenarios accurately. The
conformal and null methods require $n=2$ because they require a
regular conformal spacetime metric, and this requires the conformal
factor $\Omega\sim R^{-2}$ as $\scri$ is approached. Keeping this
asymptotic behaviour, one may be able to improve their accuracy at
large $R$ by using a smaller value of $n$ out to some very large value
of $R$ and then switching to $n=2$ to reach $\scri$.

b) In 3D implementations of compactification at $i^0$
\cite{Garfinkle,Pretorius}, Cartesian spatial coordinates are
compactified as $X^i=\tan x^i$, which corresponds to $n=2$
compactification in our notation. Our results suggest a modification
of this approach in which the slices are made asymptotically null up
to large finite radius, perhaps best with $n=1$. The resulting code
would represent outgoing radiation correctly out to this large finite
radius. Outside this region the slice could still reach $i^0$, for
example by switching to $n=2$ (the traditional value). 

c) Fixed mesh refinement \cite{MayaFMR,collapseFMR,Miller}, in 3D
typically using Cartesian grids, is also widely used to push the outer
boundary further out. If this is done with nested boxes with equal (or
comparable) numbers of grid points in each box, one automatically has
approximately constant angular resolution, and radial resolution
$\Delta R\sim R$, which corresponds to $n=1$ {\em stretching} in our
notation. Our results suggest combining the nested boxes approach with
$n=1$ {\em slicing} (and a global time step), so that outgoing waves
are resolved all the way out to the outer boundary, which can then be put
at arbitrarily large radius.


\appendix


\section{Closed form coordinate transformations}
\label{appendix:closedform}


We give closed-form functions $R(r)$ and $F(R)$ which have the
required asymptotic behaviour (\ref{RprimeLn}) and
(\ref{FprimeSchwarz}). For the radial stretch we choose
\begin{equation}
R(r)=r\left(1+{r^2\over l^2}\right)^{n\over 2(1-n)}, \qquad 
l=\left({1\over 1-n}\right)^{1/n}L
\end{equation}
for the generic case $0<n<1$, and 
\begin{equation}
R(r)=L\sinh{r\over L}
\end{equation}
for $n=1$. The range of $r$ is $0\le r<\infty$. For the
compactification we choose
\begin{equation}\label{compactification}
R(r)=r\left(1-{r^2\over l^2}\right)^{1\over 1-n}, \qquad
l=\left({2\over n-1}\right)^{1/n}L
\end{equation}
for $1<n\le 2$. The range of $r$ is now $0\le r< l$. 

The expression for $F(R)$ can be assembled from the functions
\begin{equation}
G_n(R,L)={L^n\over 1-n}\left(R^2+L^2\right)^{{1-n\over 2}}-{L\over
  {1-n}}
\end{equation}
for $0\le n \le 2$ with $n\ne 1$ and
\begin{equation}
G_1(R,L)={L\over 2}\ln\left(1+{R^2\over L^2}\right)
\end{equation}
for the special value $n=1$. They obey
\begin{equation}
G_n'(R,L)=\left({R\over L}\right)^{-n}+O(R^{-n-2}).
\end{equation}
for all $n$ including $n=1$. In particular,
\begin{equation}
G_0'(R,L)=1-{L^2\over 2 R^2}+O(R^{-4}), 
\end{equation}
and the $R^{-2}$ term in $G_0'$ has to be taken into account in the
$n=2$ slicing.  It is straightforward to construct $F(R)$ from these
blocks. We only give an example here. For $0<n<2$ with $m=q=0$, we can
use
\begin{equation}
F(R)=G_0(R,L_0)-\left({L_n\over L}\right)^n G_n(R,L_n). 
\end{equation}
Note that $L_0$ and $L_n$ can differ from each other and from $L$ (the
constant introduced in $R(r)$) as long as the corresponding amplitude
is adjusted. For $n\ge 1$ with $m>0$ and again for $n=2$, additional
terms have to be added.  Similarly, to achieve a switchover from $n$
to $n_\infty<n$ at a given large $R$ we only need to add 
$G_{n_\infty}$ with suitable amplitude and scale.


\section{Wave equation and boundary conditions}
\label{appendix:waveequation}


We write the massless wave equation
\begin{equation}
\nabla_a\nabla^a \phi = 0
\end{equation}
in a $3+1$ form similar to the ADM form of the Einstein equations by
defining
\begin{equation}
\Pi \equiv - \Lie_n \phi = - n^a\nabla_n \phi, 
\end{equation}
where $n_a$ is the future-pointing unit normal on the slices of
constant $t$. With $t^a\equiv(\partial/\partial t)^a\equiv \alpha
n^a+\beta^a$ this gives
\begin{eqnarray}
\label{wave1}
\Lie_t \phi &=& \Lie_\beta\phi -
\alpha \Pi, \\
\label{wave2}
\Lie_t \Pi &=& \Lie_\beta \Pi  - \alpha D_aD^a \phi - \alpha a^bD_b \phi
+ \alpha K\Pi.
\end{eqnarray}
Here $D_a$ is the covariant derivative associated with the
three-metric $\gamma_{ab}$ on each slice, and $a^b=D^b\ln\alpha$ is
the acceleration of the $n^a$ observers.

At the outer boundary the MDBC corresponding to no incoming
radiation \cite{GKO} would be,
\begin{equation}
\Pi-m^a D_a\phi=0,
\end{equation}
where $m^a$ is the outward-pointing unit vector normal to the
boundary within each constant $t$ slice. In Minkowski this is
equivalent to $\phi=f(R-T)$. By contrast an outgoing spherical wave in
Minkowski is $\phi=f(R-T)/R$. We can achieve this boundary condition
by adding a lower-order term to the MDBC, namely 
\begin{equation}
\Pi-m^a D_a\phi+Q\phi=0, \quad 
Q\equiv {\alpha-\beta^r\sqrt{\gamma_{rr}}\over \sqrt{\gamma_{\theta\theta}}}.
\end{equation}
Note that a MDBC is defined to lead to a non-increasing energy in the
frozen coefficient, principal part only approximation, so that this
modified BC is still a MDBC.

Following
\cite{boundshifted}, we discretise this to 2nd-order accuracy as
\begin{eqnarray}
\Pi_{N}  - \sqrt{g^{rr}}D_0\phi_N +Q_N\phi_N&=& 0, \label{boundary}\\
h^3D_{-}^3\Pi_{N+1} &=& 0,
\end{eqnarray}
where $N$ is the grid point at $r=r_{\rm max}$ and $N+1$ is a ghost
point. For 4-th order accuracy we use
\begin{eqnarray}
\Pi_{N}  - \sqrt{g^{rr}}D^{(1)}\phi_{N} +Q_N\phi_N&=& 0,\\
h^5D_+^5\phi_{N+2}&=&0,\\
h^4D_+^4\Pi_{N+1}&=&0,\\
h^4D_+^4\Pi_{N+2}&=&0,
\end{eqnarray}
where $D^{(1)}$ is defined by
\begin{equation}
D^{(1)}\equiv D_0\Bigg(1-\frac{h^2}{6}D_+D_-\Bigg).
\end{equation}
At the black hole-excision boundary, or where the outer boundary is
null or spacelike, no BCs are required for the
continuum equations. As numerical BCs we then use
extrapolation for all ghost points including $\phi_{N+1}$, that is
\begin{equation}
h^5 D_+^5\phi_{N+1}=0
\end{equation}
in the 4-th order accurate case, and $D_+^3$ in the 2nd-order accurate case.

Our artificial dissipation operators are $-\sigma h^3 (D_+D_-)^2$ at
2nd-order accuracy and $\sigma h^5(D_+D_-)^3$ at 4th order. We use
extrapolation to populate the additional ghost points required at
boundaries.




\end{document}